\documentclass[twocolumn,aps,prl,a4paper,10pt]{revtex4-1}
\usepackage{amsmath,amssymb,graphicx}

\begin{document}

\newlength{\figwidth}
\setlength{\figwidth}{\columnwidth}
\renewcommand{\vec}[1]{\mathbf{#1}}

\title{Isomorphs in the phase diagram of a model liquid without inverse power law repulsion}
\author{Arno A. Veldhorst, Lasse B{\o}hling, Jeppe C. Dyre, and Thomas B. Schr{\o}der}
\affiliation{DNRF Center ``Glass and Time'', IMFUFA, Dept. of Sciences, Roskilde University, P.O. Box 260, DK-4000 Roskilde, Denmark}
\date{\today}

\begin{abstract}
	It is demonstrated by molecular dynamics simulations that liquids interacting via the Buckingham potential are strongly correlating, i.e., have regions of their phase diagram where constant-volume equilibrium fluctuations in the virial and potential energy are strongly correlated. A binary Buckingham liquid is cooled to a viscous phase and shown to have isomorphs, which are curves in the phase diagram along which structure and dynamics in appropriate units are invariant to a good approximation. To test this, the radial distribution function, and both the incoherent and coherent intermediate scattering function are calculated. The results are shown to reflect a hidden scale invariance; despite its exponential repulsion the Buckingham potential is well approximated by an inverse power-law plus a linear term in the region of the first peak of the radial distribution function. As a consequence the dynamics of the viscous Buckingham liquid is mimicked by a corresponding model with purely repulsive inverse-power-law interactions. The results presented here closely resemble earlier results for Lennard-Jones type liquids,  demonstrating that the existence of strong correlations and isomorphs does \emph{not} depend critically on the mathematical form of the repulsion being an inverse power law.
\end{abstract}

\maketitle

\section{Introduction}
Recently a series of papers has been published concerning so-called strongly correlating liquids and their physical properties~\cite{Pedersen2008,paper1,paper2,paper3,paper4,paper5}. Liquids that exhibit these strong correlations have simpler thermodynamic, structural, and dynamical properties than liquids in general. A strongly correlating liquid is identified by looking at the correlation coefficient of the equilibrium fluctuations of the potential energy $U(\vec{r}_1,..., \vec{r}_N)$ and virial $W(\vec{r}_1,..., \vec{r}_N) \equiv -1/3\sum_i \vec{r}_i \cdot \vec{\nabla}_{\vec{r}_i} U(\vec{r}_1,..., \vec{r}_N)$~\cite{AllenTildesley} at constant volume:
\begin{equation}
	R = \frac{\left< \Delta W \Delta U \right>}
	         {\sqrt{\left<(\Delta W)^2\right> \left<(\Delta U)^2\right>}}\,.
	\label{eq:R}
\end{equation}
Here brackets denote averages in the NVT ensemble (fixed particle number, volume, and temperature), $\Delta$ denotes the difference from the average. The virial $W$ gives the configurational part of the pressure~\cite{AllenTildesley},
\begin{equation}
	pV = N k_B T(\vec{p}_1, \ldots, \vec{p}_N) + W(\vec{r}_1, \ldots, \vec{r}_N).
	\label{eq:pressure}
\end{equation}
Strongly correlating liquids are defined~\cite{Pedersen2008} as liquids that have $R\ge 0.9$.

The origin of strong $WU$ correlations was investigated in detail in Refs.~\cite{paper2,paper3} for systems interacting via the Lennard-Jones (LJ) potential:
\begin{equation}
	\upsilon(r) = 4\epsilon\left[\left(\frac{\sigma}{r}\right)^{12}
					- \left(\frac{\sigma}{r}\right)^{6}\right].
	\label{eq:gLJ}
\end{equation}
The fluctuations of $W$ and $U$ are dominated by fluctuations of pair distances within the first neighbor shell, where the LJ potential is well approximated by an extended power Law (eIPL), defined as an inverse power law (IPL) plus a linear term~\cite{paper2}:
\begin{equation}
	  \upsilon_{eIPL}(r) = Ar^{-n} + B + Cr.
	  \label{eq:eIPL}
\end{equation}
The IPL term gives perfect $UW$ correlations, whereas the linear term contributes little to the fluctuations at constant volume: when one pair distance increases, others decrease, keeping the contributions from the linear term almost constant (this cancellation is exact in one dimension). The consequence is that LJ systems inherit some of the scaling properties of the IPL potential --  they have a ``hidden scale invariance''~\cite{paper3,Schroder2009}. Prominent among the properties of strongly correlating liquids is that they have ``isomorphs'', i.e., curves in the phase diagram along which structure, dynamics, and some thermodynamical properties are invariant in appropriate units~\cite{paper4,paper5}. The physics of strongly correlating liquids was briefly reviewed recently in Ref.~\cite{Pedersen2011}.

Since the LJ systems consists of two IPL terms, it is perhaps tempting to assume that a repulsive (inverse) power law is necessary for the hidden scale invariance described above. In the present paper we use the modified Buckingham (exp-six) pair potential to show that this is \emph{not} the case. The Buckingham potential was first derived by Slater from first-principle calculations of the force between helium atoms~\cite{Slater1928}. Buckingham later used this form of the potential to calculate the equation of state for different noble gases~\cite{Buckingham1938}. The Buckingham potential has an exponential repulsive term, while the attractive part is given by a power law~\cite{Young1981,Koci2007}:
\begin{equation}
	\upsilon(r) = \epsilon\left(
		 \tfrac{6}{\alpha-6}      \exp\left[\alpha\left(1-\tfrac{r}{r_m}\right)\right]
		-\tfrac{\alpha}{\alpha-6} \left(\tfrac{r_m}{r}\right)^{\scriptscriptstyle{6}}\right).
	\label{eq:Buck}
\end{equation}
Here $\epsilon$ is the depth of the potential well and $r_m$ specifies the position of the potential minimum. The parameter $\alpha$ determines the shape of the potential well. The Buckingham potential is better able to reproduce experimental data of inert gasses than the LJ potential~\cite{Mason1954,Kilpatrick1955,Abrahamson1963}, but is also computationally more expensive (unless look-up tables are utilized~\cite{AllenTildesley}).

All simulation data in this paper were obtained from molecular dynamics in the NVT ensemble. The samples contained 1000 particles. The simulations were set up by instant cooling from a high temperature state point followed by an equilibration period, to ensure the simulations were independent from each other. The simulations were performed with the RUMD molecular dynamics package~\cite{RUMD}, which is optimized for doing computations on state-of-the-art GPU hardware.

\section{Correlations in single-component Buckingham liquids}

\begin{figure}
	\includegraphics[width=\figwidth]{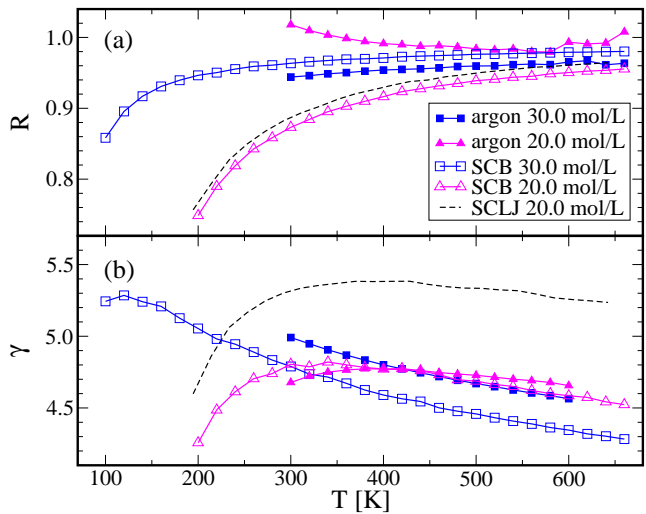}
	\caption{(a) The correlation coefficient, $R$, as a function of temperature on isochores for single-component Buckingham (SCB), single-component Lennard-Jones (SCLJ), and argon. For SCB and SCLJ  argon values were used for all potential parameters~\cite{Mason1954}, and  $R$ was calculated directly from Eq.~(\ref{eq:R}). For argon  $R$ was calculated from experimental data~\cite{NIST} as described in Refs.~\cite{Pedersen2008,paper2}. The correlations are strongest for state points with both high density and high temperature, and the difference between the Buckingham and the LJ potential is small.  The correlation coefficient $R>1$ for low-temperature $20.0~\text{mol/L}$ argon is of course unphysical and either caused by an uncertainty in the experimental data or the approximations applied in the calculation of $R$ (see Refs.~\cite{paper2,Pedersen2008suplement} for details). (b) The value of $\gamma$ (Eq.~(\ref{eq:gamma1})) plotted versus temperature for the same systems as in (a). For argon  $\gamma$ was calculated from experimental data~\cite{NIST} using Eq.~(\ref{eq:gamma_argon}). $\gamma$ decreases slowly for increasing temperatures, except when the correlation coefficient is low ($R\lesssim 0.9$).}
	\label{fig:correlations}
\end{figure}

To compare the simulations with experiments~\cite{NIST}, argon parameters from Ref.~\cite{Mason1954} were used; $\alpha = 14.0$, $r_m = 0.3866~\text{nm}$, $\epsilon/k_B = 123.2~\text{K}$. As can be seen in Fig.~\ref{fig:correlations}(a), the single-component Buckingham (SCB) liquid is strongly correlating ( $R\ge 0.9$) in parts of the phase diagram, particularly at high densities and/or temperatures. The correlation coefficients (Eq.~(\ref{eq:R})) of the Buckingham systems are very similar to those of argon and the LJ system (dotted line in Fig.~\ref{fig:correlations}). This is a first indication that the actual functional form of the repulsive part of the potential does not have to be an inverse power law in order for a system to exhibit strong $WU$ correlations.

Another interesting property of the fluctuations is the quantity $\gamma$ defined~\cite{paper3,paper4} as
\begin{equation}
	\gamma = \frac {\left<\Delta W \Delta U\right>} {\left<(\Delta U)^2\right>}.
	\label{eq:gamma1}
\end{equation}
When a system is strongly correlating ($R$ is close to one), $\Delta W \approx \gamma \Delta U$. For IPL potentials $\gamma$ is constant and equal to $n/3$ and $R=1$. For non-IPL potentials, however, $R<1$ and $\gamma$ may change with temperature and density as seen in Fig.~\ref{fig:correlations}(b)~\cite{Pedersen2008,Coslovich2009}. Especially for $R<0.9$, we find $\gamma$ changing rapidly. The curves are similar for the $20.0~\text{mol/L}$ SCB and SCLJ systems, except for a vertical offset.

Fluctuations in $U$ and $W$ are of course only directly accessible in simulations. For experimental systems one must revert to the use of thermodynamic quantities that reflect the fluctuations in $U$ and $W$. For instance, the configurational part of the pressure coefficient $\beta^{ex}_V = (\partial(W/V)/\partial T)_V$ and the configurational part of the isochoric specific heat per unit volume $c^{ex}_V =(\partial(U/V)/\partial T)_V$ can be used to to calculate $\gamma$ for argon as follows~\cite{paper2,paper4,Pedersen2008}:
\begin{equation}
	\gamma = \frac {\beta^{ex}_V} {c^{ex}_V}
	\label{eq:gamma_argon}
\end{equation}
The values of $\gamma$ for argon obtained in this way are  plotted in Fig.~\ref{fig:correlations}(b), and the agreement with the Buckingham systems is good. This confirms that the Buckingham potential produces more accurate predictions of experimental argon data than the LJ potential.

Interestingly, low density argon has a higher correlation coefficient than high density argon. This  is the opposite of what is found for the Buckingham and the LJ potentials. Furthermore, the buckingham data are in better agreement with the argon data at low density than at high density. At the present we do not have any explanation for these observations.

\section{Isomorphs in binary Buckingham mixtures}
Strongly correlating liquids are predicted to have isomorphs, which are curves in the phase diagram along which structure, dynamics, and some thermodynamical properties are invariant in appropriate reduced units~\cite{paper4,paper5}. Introducing reduced coordinates as $\tilde{\vec{r}}_i = \rho^{1/3}\vec{r}_i$, two state points (1) and (2) are defined to be isomorphic if pairs of microscopic configurations with same reduced coordinates ($\tilde{\vec{r}}_i^{(1)} = \tilde{\vec{r}}_i^{(2)}$)  have proportional configurational Boltzmann weights:
\begin{equation}
	       e^{-U(\vec{r}_1^{(1)},...,\vec{r}_N^{(1)})/k_B T_1} =
	C_{12} e^{-U(\vec{r}_1^{(2)},...,\vec{r}_N^{(2)})/k_B T_2} \label{Eq:IsomorphDef}.
\end{equation}
Here the constant $C_{12}$ depends only on the two state points and Eq.~(\ref{Eq:IsomorphDef}) is required to hold to a good approximation for all physically relevant configurations~\cite{paper5}. An isomorph is a curve in the state diagram for which all points are isomorphic (an isomorph is a mathematical equivalence class of isomorphic state points). The isomorphic invariance of structure, dynamics, and some thermodynamical properties -- all in reduced units -- can be derived directly from Eq.~(\ref{Eq:IsomorphDef})~\cite{paper4}. Only IPL liquids have exact isomorphs, but it has been shown that all strongly correlating liquids have isomorphs to a good approximation (Appendix~A of Ref.~\cite{paper4}).

Among the thermodynamical properties that are isomorphic invariant is the excess entropy, $S^{ex}\equiv S - S_{ideal}$, where  $S_{ideal}$ is the entropy of an ideal gas at the same temperature and density.  In the following, isomorphic state points were generated by utilizing that the quantity $\gamma$ in Eq.~(\ref{eq:gamma1}) can be used to change density and temperature while keeping the excess entropy constant~\cite{paper4,paper5}:
\begin{equation}
	\gamma = \left(\frac{\partial \ln T}{\partial \ln \rho}\right)_{S_{ex}}\,.
	\label{eq:gamma2}
\end{equation}
By choosing the density of a new isomorphic state point close to the density of the previous isomorphic state point, the temperature of the new state point can be calculated from the fluctuations by combining Eq.~(\ref{eq:gamma1}) and Eq.~(\ref{eq:gamma2})~\cite{paper4}. In this way a set of isomorphic points can be obtained from one initial state point.

The predicted isomorphic invariance of the dynamics is most striking in viscous liquids, where the dynamics in general depend strongly on temperature and density. To demonstrate that a systems interacting via the Buckingham potential have isomorphs, we study what we term a Kob-Andersen binary Buckingham (KABB) mixture with potential parameters being the same as for the original Kob-Andersen binary LJ (KABLJ) mixture~\cite{Kob1994}: $\epsilon_{AA}=1.0$, $r_{m,AA}=\sqrt[6]{2}$, $\epsilon_{AB}=1.5$, $r_{m,AB}=0.8\sqrt[6]{2}$, $\epsilon_{BB}=0.5$, $r_{m,BB}=0.88\sqrt[6]{2}$. A 4:1 mixture (A:B) was used with $\alpha = 14.5$.  The potentials were truncated and shifted at $r_{ij}^{cut}=2.5 r_{m,ij}/\sqrt[6]{2}$.

\begin{figure}
	\includegraphics[width=\figwidth]{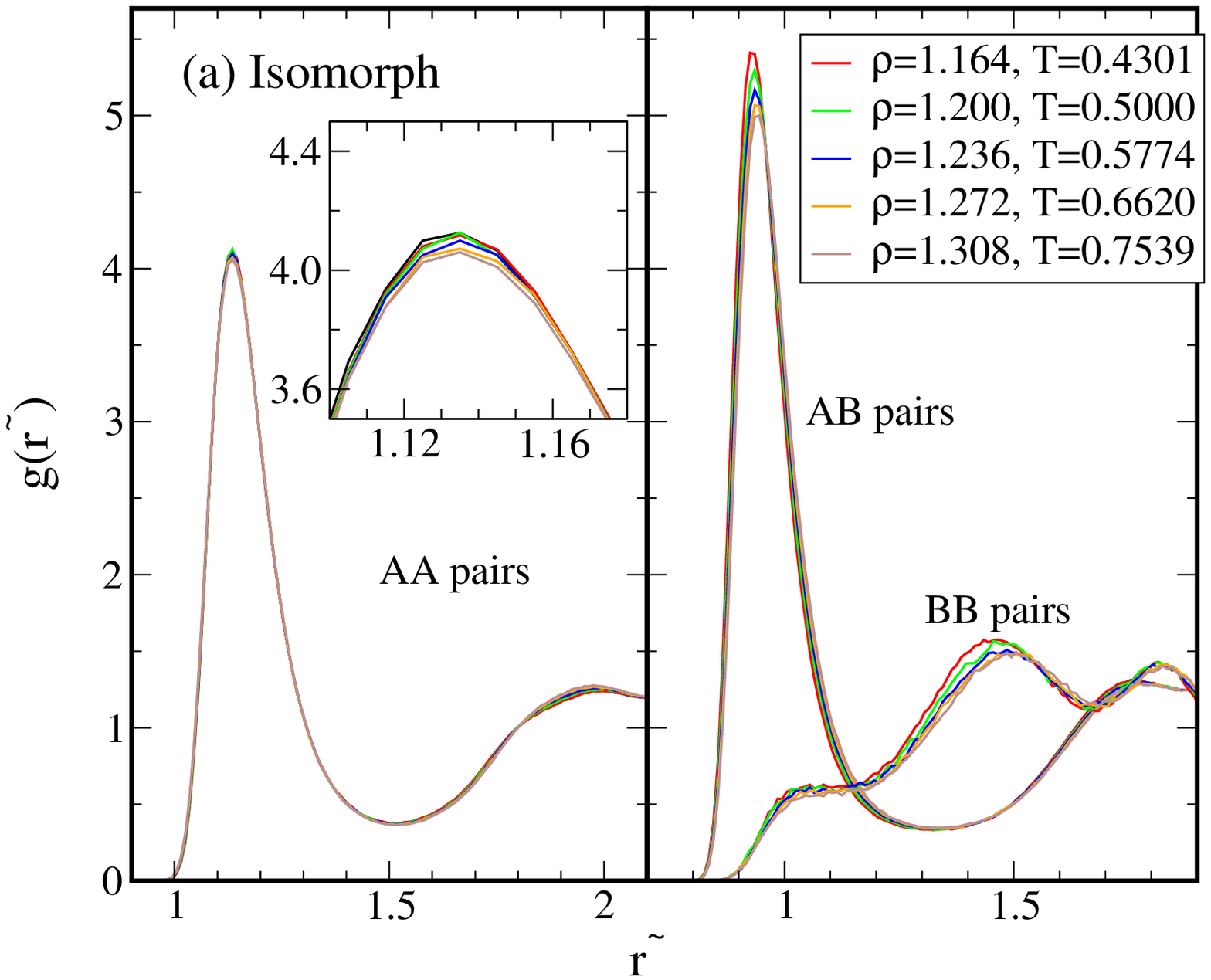}
	\includegraphics[width=\figwidth]{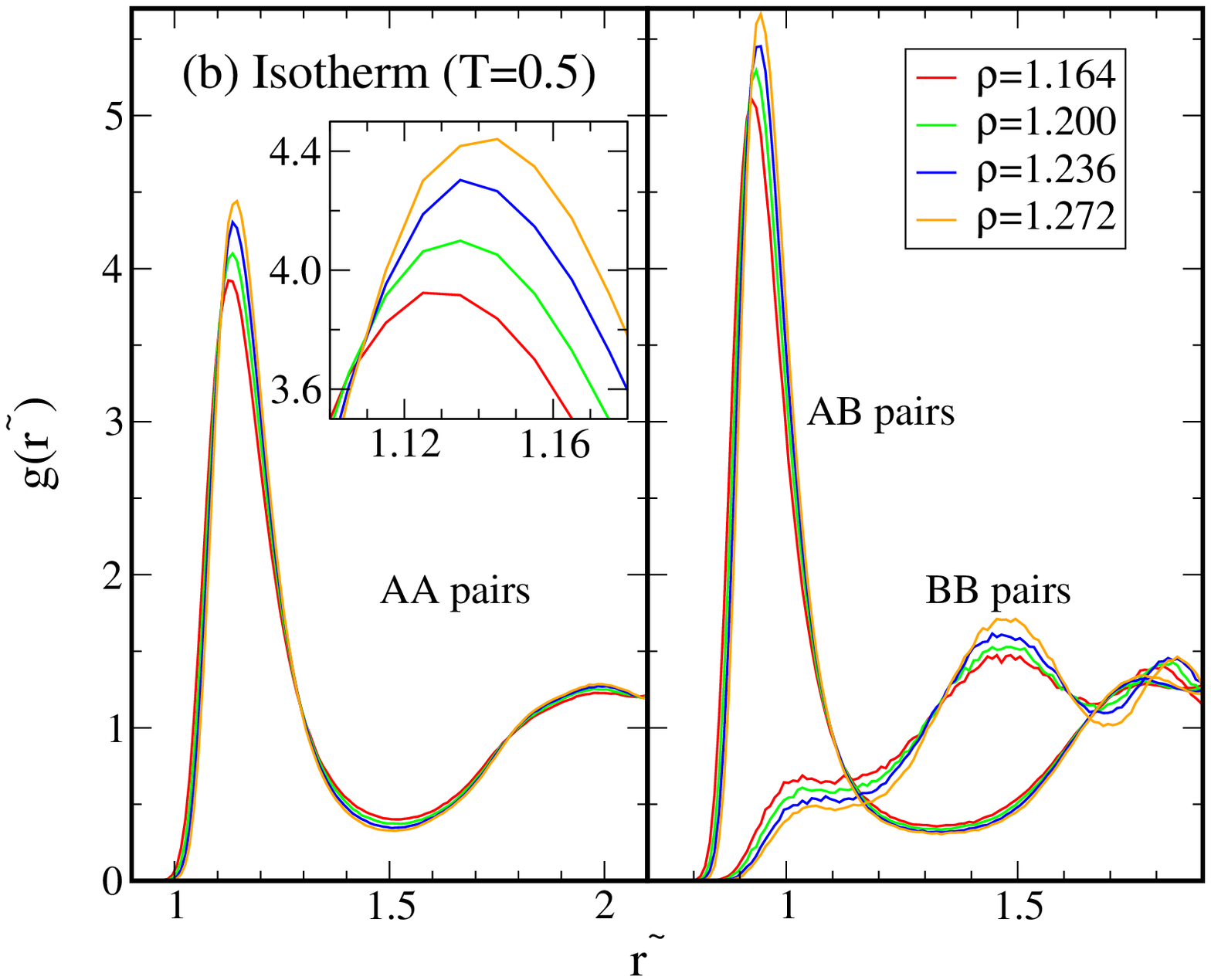}
	\caption{The radial distribution functions $g(r)$ for simulations of the KABB mixture. Both graphs are in reduced units where $\tilde{r}=\rho^{1/3}r$. (a) $g(\tilde{r})$ for isomorphic state points and the three different particle combinations. The structure is invariant on the isomorph for the AA particle pairs, but for the AB and BB pairs the structure is less invariant. (b) $g(\tilde{r})$ for isothermal state points of smaller density variation. The structure is not invariant on the isotherm for any of the particle pairs.}
	\label{fig:rdf}
\end{figure}

One of the predicted invariants on an isomorph is the structure of the system. To test this prediction, the radial distribution function in reduced coordinates $g(\tilde{r}) = g(\rho^{1/3}r)$ was plotted for isomorphic state points (Fig.~\ref{fig:rdf}(a)). The structure is invariant for the large (A)  particle pair correlation function to a very good approximation. For the AB and BB pairs the structure is less invariant. However, when a comparison is made with Fig.~\ref{fig:rdf}(b), it is clear that $g(\tilde{r})$ for the AB and BB pairs is still more invariant on an isomorph than on an isotherm (note that the density variation on the isomorph is larger than on the isotherm). This situation is similar to what is found for the KABLJ system~\cite{paper4}.

\begin{figure}
	\includegraphics[width=\figwidth]{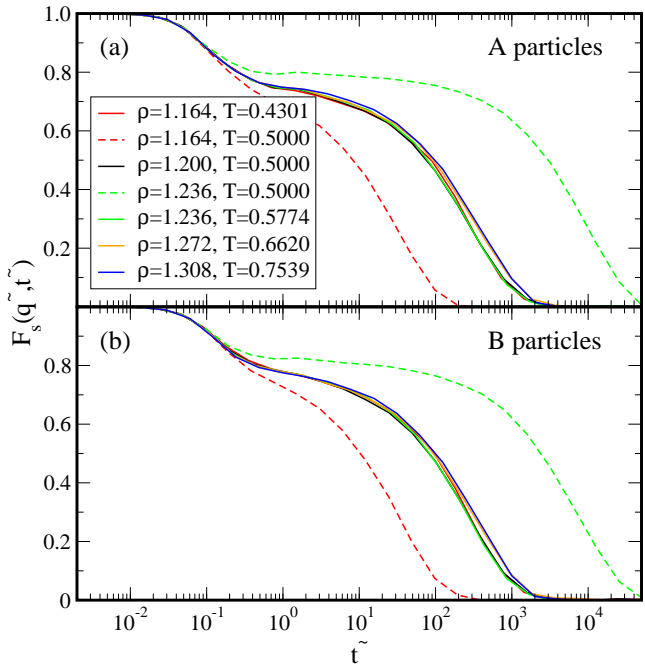}
	\caption{Incoherent intermediate scattering function for the A (a) and B (b) particles of the KABB system. The time is given  in reduced units ($\tilde{t} = \rho^{1/3}T^{1/2}t$) and the q-vector is kept constant in reduced units: $q_A = 7.25(\rho/1.2)^{1/3}$ and $q_B = 5.5(\rho/1.2)^{1/3}$. The solid lines represent isomorphic state points, while dashed lines show isothermal density changes for comparison. The dynamics are to a good approximation invariant on an isomorph when expressed in reduced units, especially when compared to the isotherm. In contrast to $g(\tilde{r})$, this holds for both the A and the B particles.}
	\label{fig:isom_Fs}
\end{figure}

To investigate the dynamics of the systems, the incoherent intermediate scattering function $F_s(q,t)$ is plotted in reduced units in Fig.~\ref{fig:isom_Fs}(a) and Fig.~\ref{fig:isom_Fs}(b). The presence of a plateau in $F_s$ shows that the system is in a viscous state, where the dynamics are highly state point dependent. The large difference in $F_s$ for the two isothermal state points confirms this (dashed lines). For the isomorph all $F_s$ data collapse more or less onto the same curve, showing that the dynamics are indeed invariant to a good approximation on an isomorph. In contrast to the radial distribution functions, the invariance holds well for both types of particles.

\begin{figure}
	\includegraphics[width=\figwidth]{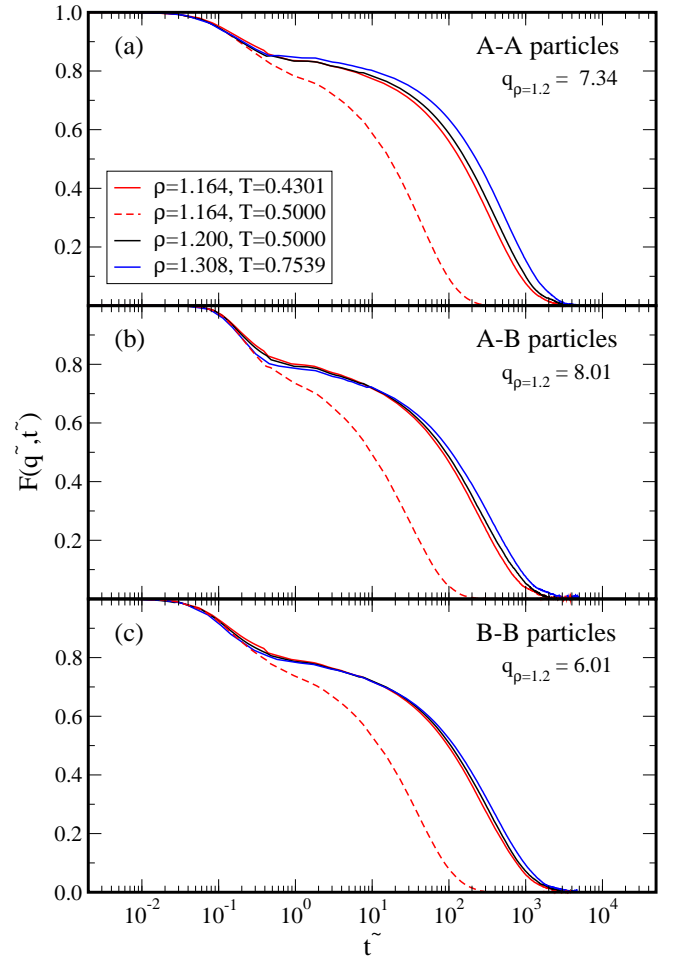}
	\caption{Coherent intermediate scattering function for different particle pairs on an isomorph in reduced units ($\tilde{t} = \rho^{1/3}T^{1/2}t$). The solid lines represent isomorphic state points, while dotted lines show isothermal density changes for comparison. Again, the q-vector is kept constant in reduced units: $q_{AA} = 7.34(\rho/1.2)^{1/3}$, $q_{AB} = 8.01(\rho/1.2)^{1/3}$ and $q_{BB} = 6.01(\rho/1.2)^{1/3}$. Also $F(\tilde{q},\tilde{t})$ is invariant on the isomorph. Contrary to what is seen for $g(\tilde{r})$, the invariance holds better for the AB and BB parts.}
	\label{fig:isom_F}
\end{figure}

To investigate the invariance in dynamics further, the coherent intermediate scattering function was calculated (Fig.~\ref{fig:isom_F}). The coherent intermediate scattering function was calculated from the spatial transform of the number density $\rho(q)$~\cite{AllenTildesley}. In order to obtain good results, it is necessary to average over time scales that are 10-15 times longer than what is usual for the intermediate scattering function. This is the reason that there are less state points shown for the coherent-, than for the incoherent intermediate scattering function. The data confirm that the dynamics are invariant on the isomorph, especially when compared to the isothermal density change (dashed lines). However, the invariance seems to hold slightly better for the AB and BB parts, which is the opposite of what is seen for the structural invariance.

\begin{figure}
	\includegraphics[width=\figwidth]{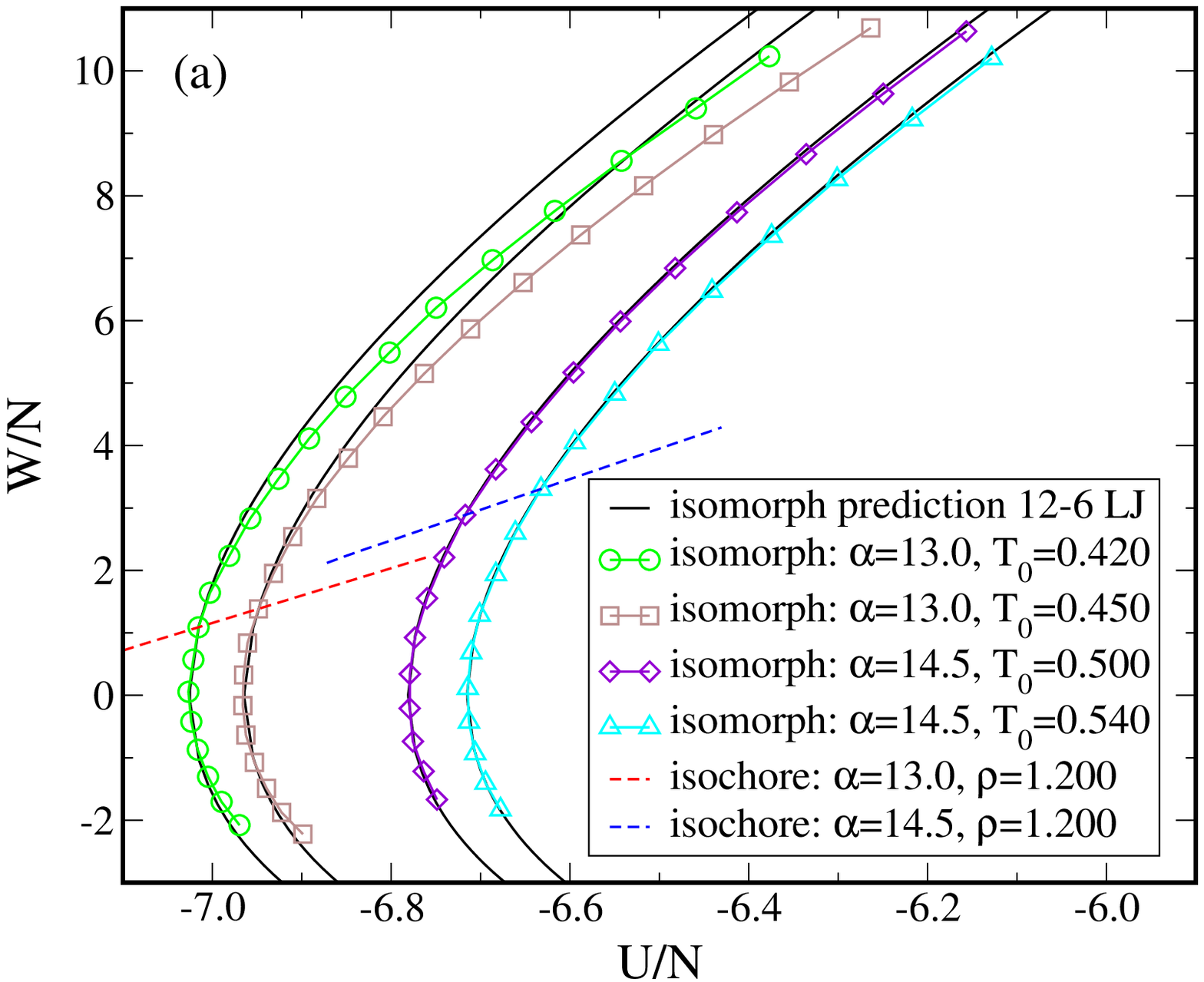}
	\includegraphics[width=\figwidth]{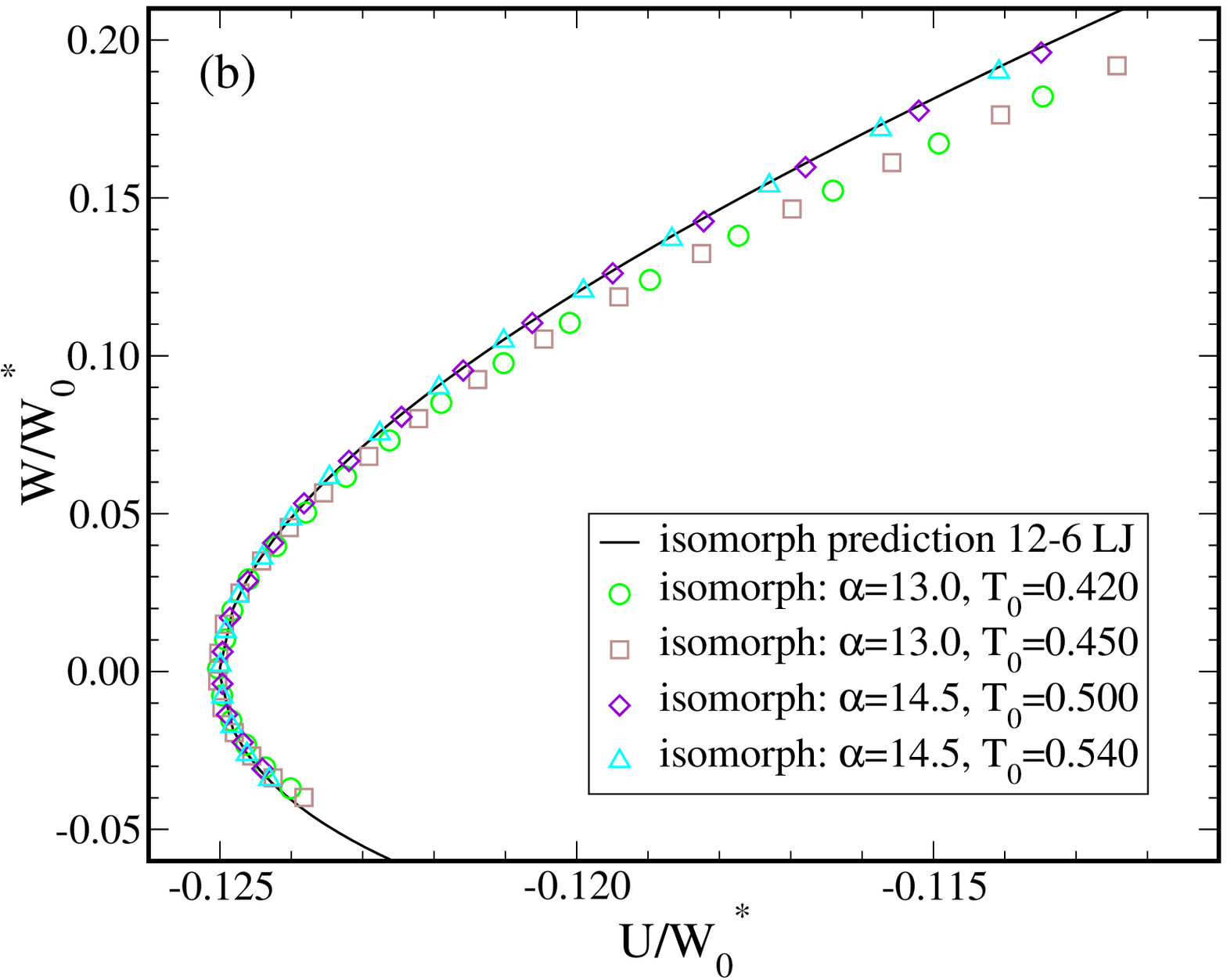}
	\caption{(a) Plot of the potential energy per particle versus virial per particle for the KABB system. The solid lines are the predictions of the isomorph shape for the 12-6~LJ potential~\cite{paper5}. For these predictions the initial state points with $\rho = 1.2$ were used as reference point. Since a new value of $\gamma$ was calculated for every state point, $\gamma$ is not constant on the isomorphs, but changes approximately 10\% along the isomorphs. The shape of the KABB isomorph agrees very well with the predicted shape for the 12-6~LJ potential for $\alpha = 14.5$. For $\alpha = 13.0$, the shape is different. (b) The same data now scaled with $W^*_0$ defined as the virial at $U=0$~\cite{paper5}. The isomorphs scale onto each other, forming a so called master isomorph for each value of $\alpha$.}
	\label{fig:isom_UW}
\end{figure}

For systems described by a generalized LJ potential consisting of two IPL terms, the invariance of the structure leads to a prediction for the shape of an isomorph when plotted in the $U$-$W$ plane~\cite{paper5} (generalized LJ potentials are a sum of inverse power laws). Since the repulsive term in the Buckingham potential is described by an exponential function, it is not possible to derive an exact equation that describes the isomorph in terms of $U$ and $W$. Figure~\ref{fig:isom_UW}(a) shows that isomorphs for the KABB system agree well with the prediction for the 12-6~LJ system if $\alpha = 14.5$ (this value of $\alpha$ was chosen to demonstrate this feature). For $\alpha = 13.0$, there is a significant difference with the predicted shape at higher density and temperature. Figure~\ref{fig:isom_UW}(b) shows the isomorphs for both values of $\alpha$ after scaling $U$ and $W$ by the same isomorph-dependent factor, demonstrating the existence of a master isomorph~\cite{paper5}. This shows that master isomorphs exist not only in generalized LJ systems where they can be justified from analytical arguments~\cite{paper5}.

\section{The inverse-power-law (IPL) approximation}
\begin{figure}
	\includegraphics[width=\figwidth]{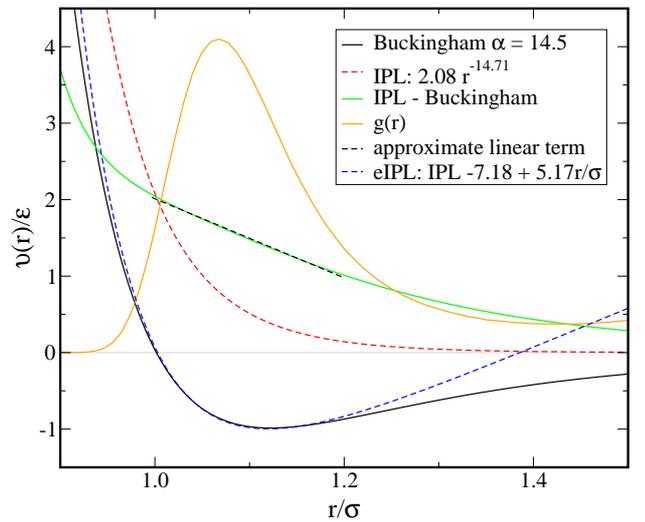}
	\caption{The figure shows how the Buckingham potential ($\alpha=14.5$) can be approximated by an extended IPL potential (eIPL). The red dotted line is the IPL approximation obtained using the parameters obtained below in Fig.~\ref{fig:ipl_param}. The difference of the IPL approximation and the Buckingham potential (dashed green line) is more or less linear in the first peak of $g(r)$. By subtracting this linear term from the IPL term the eIPL approximation is found (dashed blue line).}
	\label{fig:eipl}
\end{figure}

As mentioned in the introduction, a generic explanation~\cite{paper2,paper3,paper4} for the existence of strong correlations and isomorphs in non-IPL systems, is the fact that some pair potentials can be well approximated by an eIPL (Eq.~\ref{eq:eIPL}) as shown in Fig.~\ref{fig:eipl}. Putting this explanation to a test, it was recently demonstrated that structure and dynamics of the KABLJ system can be reproduced by a purely repulsive IPL system even in the viscous phase~\cite{Pedersen2010}. In the following we demonstrate that this procedure works also for the KABB system, despite its non-IPL repulsion. 

Following Pedersen et al.~\cite{Pedersen2010}, we assume that the Kob-Andersen IPL (KABIPL) system used to approximate the KABB system has the form
\begin{equation}
	\upsilon^{IPL}(r) = A \epsilon_{ij} \left(\frac{\sigma_{ij}}{r_{ij}}\right)^n
	\label{eq:u_ipl}
\end{equation}
where the parameters $\epsilon_{ij}$ and $\sigma_{ij}$ are the Kob-Andersen parameters for the different types of particles and the constants $A$ and $n$ are independent of particle type.

\begin{figure}
	\includegraphics[width=\figwidth]{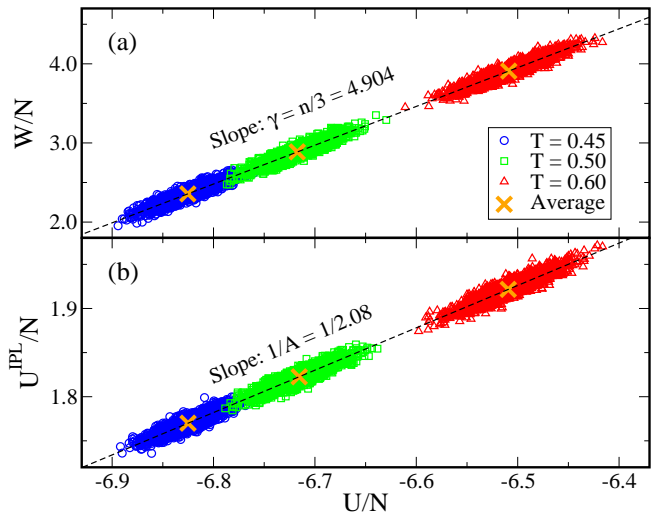}
	\caption{Three isochoric state points were used to obtain the parameters for the IPL potential. (a) The value of $n$ was determined by linear regression to  the mean values of the virial and the potential energy(marked by yellow crosses. (b) The method used to find the value of $A$ of Eq. (\ref{eq:U_ipl}). $U^{IPL} = \sum_{i>j} \epsilon_{ij} \left( \sigma_{ij}/r_{ij}\right)^n$ was calculated from configurations drawn from the KABB simulations and plotted against the energy obtained during the simulations; the value of $A$ was then obtained from the slope of the mean energies (again marked by yellow crosses).}
	\label{fig:ipl_param}
\end{figure}

For IPL liquids it is known that $W=(n/3)U$, so in principle the value of $n$ could be calculated from $\gamma$ determined from the $WU$ fluctuations (Eq.~(\ref{eq:gamma2})). For non-IPL liquids however, there is a slight state point dependence of $\gamma$, so instead the slope of an isochore was used to determine $n$ (Fig.~\ref{fig:ipl_param}(a)) making use of the identity~\cite{paper4}
\begin{equation}
\gamma\,=\,\left( \frac{\partial W}{\partial U}\right)_V\,.
\end{equation}
For the Buckingham potential with $\alpha = 14.5$ we obtained $\gamma=4.904$ and $n=14.71$. This is lower than the $\gamma=5.16$ which was found for the 12-6~LJ potential~\cite{Pedersen2010}. This is also consistent with the data in Fig.~\ref{fig:correlations}(b) where the SCB system has a lower value of gamma than the SCLJ system.

\begin{figure}
	\includegraphics[width=\figwidth]{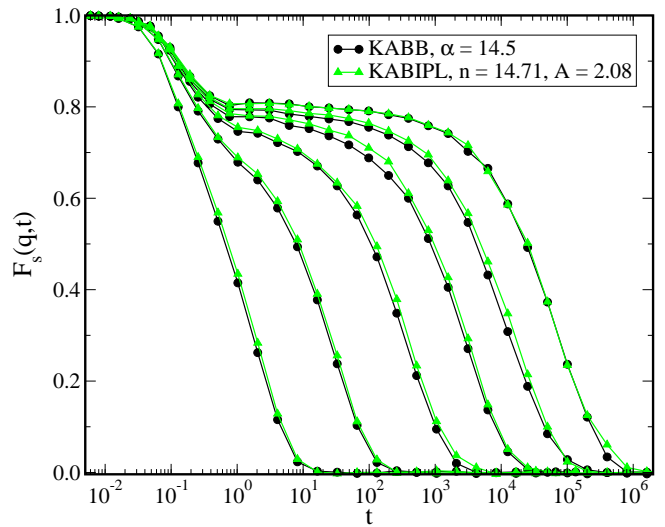}
	\caption{The incoherent intermediate scattering function $F_s(q,t)$, ($q=7.25$) of the KABB and KABIPL simulations for isochoric state points with $\rho = 1.2$, $T = 0.42, 0.44, 0.46, 0.50, 0.6, 1.0$.  The IPL potentials reproduce the dynamics of the Buckingham potential over a significant temperature range.}
	\label{fig:ipl_Fs}
\end{figure}

From Eq.~(\ref{eq:u_ipl}) it follows that the total internal energy of the IPL system can be written as
\begin{equation}
	U^{IPL} = A \sum_{i>j} \epsilon_{ij} \left(\frac{\sigma_{ij}}{r_{ij}}\right)^n.
	\label{eq:U_ipl}
\end{equation}
The scaling factor $A$ was determined from the slope of the mean values of the energies in a $U$,$U^{IPL}$  plot (Fig.~\ref{fig:ipl_param}(b)), where $U^{IPL}$ is given by Eq.~(\ref{eq:U_ipl}) evaluated on configurations from simulations of the KABB mixture~\cite{Pedersen2010}. Using these parameters, simulations of the KABIPL systems were performed and the results were compared with the results of the KABB system. In Fig.~\ref{fig:ipl_Fs} the incoherent intermediate scattering function of the two systems is plotted for comparison. The KABIPL reproduces the dynamics of the KABB system very well. It should however be noted that in spite of the good reproduction of the dynamics, the KABIPL had a stronger tendency to crystallize than the KABB system at the two lowest temperatures due to the absence of attractive forces. The good agreement shown in Fig.~\ref{fig:ipl_Fs} only holds if both systems are in the same (supercooled) state.

\begin{figure}
	\includegraphics[width=\figwidth]{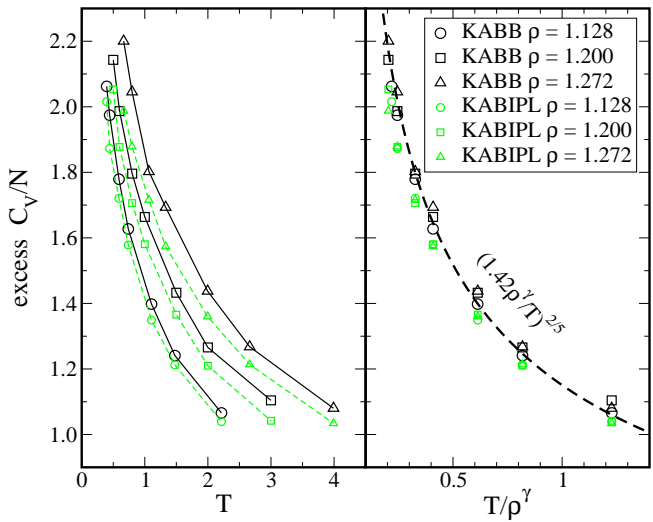}
	\caption{(a) The configurational part of the intensive isochoric specific heat $C_V^{ex}/N$ as a function of temperature. Three isochores were simulated of the KABB and the KABIPL systems. At low density the agreement between the two systems is fairly good, but for higher densities the differences become larger. (b) The same data plotted versus $T/\rho^\gamma$ where $\gamma = 4.904$. The data collapse on a single curve, which shows that density scaling works. The function $ (1.42\rho^\gamma / T)^{2/5}$ was fitted to the data (dashed line), showing that Rosenfeld-Tarazona scaling is also obeyed.}
	\label{fig:ipl_CV}
\end{figure}

From the fluctuations in the potential energy one can calculate the excess isochoric specific heat using~\cite{AllenTildesley}:
\begin{equation}
	C_V^{ex} = C_V - \frac{3}{2} N k_B = \frac {\left<(\Delta U)^2\right>} {k_BT^2}\,.
\end{equation}
In Fig.~\ref{fig:ipl_CV}(a) $C_V^{ex}$ is plotted for different isochores calculated from KABB and KABIPL simulations. The heat capacities for the two systems follow each other closely, although there is a small and systematic difference increasing with density. This is similar to what was found for the KABLJ system~\cite{Pedersen2010}, but the deviations are slightly larger for the KABB mixture. Figure~\ref{fig:ipl_CV}(b) shows that the excess heat capacity to a good approximation obeys density scaling, $C_V^{ex}/N=f(\rho^\gamma/T)$, and Rosenfeld-Tarazona scaling, $C_V^{ex}/N=g(\rho)T^{-2/5}$~\cite{Rosenfeld1998} -- again in good agreement with results for the KABLJ system~\cite{Pedersen2010}.

\section{Conclusion}
The Buckingham potential has been shown to be strongly correlating like the Lennard-Jones potential. In spite of its exponential repulsion, the Buckingham potential's dynamics and heat capacity can be closely approximated by a purely repulsive IPL system. In particular the system has good isomorphs in the phase diagram. These findings are very similar to those found for Lennard-Jones systems. We conclude that the existence of strong correlations and isomorphs is \emph{not} dependent on the repulsion being an inverse power-law.

\section*{Acknowledgments}
The centre for viscous liquid dynamics ``Glass and Time'' is sponsored by the Danish National Research Foundation (DNRF).


\end{document}